\documentclass{article}

\usepackage{arxiv}

\usepackage[utf8]{inputenc} 
\usepackage[T1]{fontenc}    
\usepackage{hyperref}       
\usepackage{url}            
\usepackage{booktabs}       
\usepackage{amsfonts}       
\usepackage{nicefrac}       
\usepackage{microtype}      
\usepackage{lipsum}		
\usepackage{graphicx}
\usepackage{natbib}
\usepackage{doi}
\usepackage{amsmath}
\usepackage{listings}
\lstdefinestyle{promptstyle}{
    basicstyle=\ttfamily\tiny,
    breaklines=true,
    frame=single,
    columns=fullflexible
}

\title{Opinion Polarization in LLM-Based Social Networks: Manipulation and Mitigation}

\author{
Ali Safarpoor Dehkordi \\
School of Computing \\
Australian National University \\
Canberra, ACT 2601, Australia \\
\texttt{Ali.SafarpoorDehkordi@anu.edu.au}
\And
Mohammad Shirzadi \\
School of Computing \\
Australian National University \\
Canberra, ACT 2601, Australia \\
\texttt{Mohammad.Shirzadi@anu.edu.au}
\And
Ahad N. Zehmakan \\
School of Computing \\
Australian National University \\
Canberra, ACT 2601, Australia \\
\texttt{ahadn.zehmakan@anu.edu.au}
}

\date{}

\hypersetup{
pdftitle={A template for the arxiv style},
pdfsubject={q-bio.NC, q-bio.QM},
pdfauthor={David S.~Hippocampus, Elias D.~Striatum},
pdfkeywords={First keyword, Second keyword, More},
}

\begin{document}
\maketitle

\begin{abstract}
How vulnerable are online social networks to adversaries who seek to amplify opinion polarization by manipulating opinions, and how difficult is it to mitigate such manipulation? Existing studies have examined this question using mathematical models of opinion dynamics. While these models offer valuable theoretical insights, they rely on simplified assumptions about interactions, message content, and opinion updates, limiting the adversarial strategies they can capture and the applicability of their findings to real-world settings. Large language model (LLM)-based simulations provide a richer alternative: agents can be assigned diverse personas, communicate through natural language, and respond to persuasive or adversarial content in a context-dependent way. This enables the study of manipulation strategies that are difficult to represent using classical mathematical models.
To the best of our knowledge, this study provides the first systematic analysis of polarization amplification and mitigation in an LLM-based simulated social network framework.
In our framework, LLM agents with diverse personas interact over a social network by exchanging natural language posts and updating their opinions accordingly. We show that even an adversary with a limited manipulation budget can considerably increase polarization. We then study two classes of defense mechanisms: reactive mitigations, which assign specific users to actively counter manipulation, and proactive interventions, which increase resistance through general mechanisms not tied to particular users. Our results show that although these mechanisms reduce the impact of adversarial attacks, they generally do not restore the network to its baseline polarization state. These findings suggest that neither approach fully overcomes the vulnerability of the network, highlighting the potential risk of such attacks. 
Our implementation is publicly available at \url{https://github.com/aSafarpoor/AA}.

\end{abstract}


\section{Introduction}

Online social networks increasingly shape information dissemination, communication, and opinion formation. Repeated interactions and exposure to like-minded users have been linked to social fragmentation, ideological extremism, reduced exposure to diverse viewpoints, and polarization~\cite{arora2022polarization}. In addition, their large-scale and interactive nature makes them vulnerable to manipulation. Malicious actors exploit these platforms to spread misinformation~\cite{vosoughi2018spread}, deploy manipulative advertising~\cite{ali2019discrimination}, conduct political interference~\cite{tucker2018social}, and encourage emotional reactions~\cite{brady2017emotion}. This raises an important question: \textit{How vulnerable are social networks to manipulators that aim to amplify polarization, and to what extent can realistic mitigation mechanisms reduce these effects?}

A large body of prior work studies opinion dynamics through mathematical models, where user opinions are represented numerically and evolve according to fixed interaction rules~\cite{gaitonde2021polarization}. These models have been widely used to analyze polarization, adversarial manipulation, and mitigation strategies~\cite{chen2021adversarial,racz2023towards,zehmakan2021majority,zehmakan2023rumors}, which improve our understanding of how network structure and update mechanisms affect the evolution of opinions in social networks.

However, the abstraction of mathematical models limits their ability to capture complex social interactions. Most approaches model influence through predefined update functions, such as numerical averaging rules~\cite{gaitonde2021polarization}, rather than through natural language communication. Consequently, such models cannot directly capture richer forms of language-based influence, such as persuasion, emotional tone, biased communication, or persona-dependent responses~\cite{chuang2024simulating,touzel2024simulation}, which are important for studying opinion manipulation in online social networks~\cite{orlando2025emergent}. Recent work has begun to address this limitation by using LLM-based simulations to study manipulation in social media environments, examining outcomes such as voting behavior~\cite{touzel2024simulation} and coordination among agents in information operations~\cite{orlando2025emergent}; however, these studies have not focused on how adversarial natural-language interactions amplify and mitigate polarization over explicit opinion states.
Recent advances in large language models (LLMs) have provided new opportunities to study social dynamics through natural-language interactions. Prior work has shown that LLM agents can reproduce polarization and coordinated interactions in social networks~\cite{chuang2024simulating,wang2025decoding,gu2025large}. 

Other studies have examined adversarial behavior in LLM-based multi-agent systems, but mainly in small collaboration settings. For example, some work studies malicious agents that introduce errors into collaborative workflows~\cite{huang2024resilience}, while others study adversarial agents that try to persuade other agents during debate~\cite{amayuelas2024multiagent,zhang2025allies}. These settings typically involve small groups of agents communicating to solve tasks, rather than users interacting over a social network where opinions evolve.
As a result, adversarial opinion manipulation in LLM-based social network simulations has not been systematically studied. In particular, it remains unclear how different adversarial strategies affect polarization and how such effects can be mitigated.

In this study, we propose a simulation framework for studying adversarial opinion manipulation and mitigation in social networks. To the best of our knowledge, this is the first work to provide a systematic analysis of how adversarial strategies influence polarization and how such effects can be mitigated in LLM-based social network simulations.
Here, users are modeled as LLM-based agents that communicate through natural language posts over a directed network, where interactions influence the opinions of neighboring users. The model captures heterogeneity among agents through differences in activeness, stubbornness, and initial opinions, which enables the study of how local interactions give rise to global polarization.

Within this framework, we define adversarial manipulators with different selection strategies, persistence levels, and manipulation budgets. We analyze how these manipulators influence opinion dynamics under different strategies and budgets. We find that even a small number of strategically placed manipulators with high activeness and high stubbornness, or full stubbornness in the persistent setting, can noticeably increase polarization and extremization across networks. This effect is consistent across different network structures and settings. These results highlight the vulnerability of social networks to targeted manipulation.

Given this vulnerability, we examine how effectively interventions can mitigate adversarial effects.
We first consider \textit{reactive} approaches,
where selected users are assigned to actively counter adversarial influence. Although these methods can reduce polarization, they do not restore the network to the no-manipulation baseline and typically require strong interventions that may be unrealistic in practice. 
We then consider \textit{proactive} interventions that modify exposure patterns, connectivity, or activity levels. These approaches are more practical and effective for improving robustness against manipulation; however, they still cannot fully counter manipulators. 
Overall, our results show that social networks remain highly vulnerable to adversarial manipulation, and the considered interventions cannot restore the system to its no-manipulation baseline.

The remainder of this paper is organized as follows: related work (Section~\ref{sec:relatedWorks}), proposed framework (Section~\ref{sec:Preliminaries}), manipulation (Section~\ref{sec:manipulation}) and mitigation (Section~\ref{sec:mitigation}) strategies, experimental evaluation (Section~\ref{sec:experiments}), and finally, the conclusion with a discussion of future work (Section~\ref{sec:conclusion}).

\section{Related Work}
\label{sec:relatedWorks}
This section reviews prior work on graph-based opinion dynamics, mitigation strategies, LLM-based social network simulations, and opinion manipulation in LLM-based social networks.

\paragraph{\textbf{Graph-Based Opinion Dynamics and Polarization}}
A large body of prior work studies opinion dynamics using graph-based models~\cite{shirzadi2025opinion,miyauchi2026survey,zareer2025survey,out2024impact}. Baumann et al.~\cite{baumann2020modeling} showed that reinforcement and homophily can drive transitions from consensus to polarization, while T\"ornberg~\cite{tornberg2022digital} argued that social networks amplify polarization by increasing interactions with people outside their local communities.
Prior studies have demonstrated that polarization is shaped by factors such as stubbornness, interaction rules, and network structure~\cite{shirzadi2025stubborn,gaitonde2021polarization}. Related work has also shown that attempts to minimize disagreement may unintentionally increase polarization~\cite{coscia2022minimizing}.
Agent-based approaches model opinion formation using heterogeneous agents, where agents differ in self-confidence, susceptibility to social influence, and community-dependent behavior~\cite{min2022multi}.
More recent work incorporates recommendation systems and captures the co-evolution of opinions and network structures~\cite{chen2025coevolution}, as well as misinformation-specific dynamics and algorithmic exposure, which can contribute to radicalization and polarization~\cite{zhang2024misinformation}.

\paragraph{\textbf{Mitigation of Polarization and Misinformation}}
Several studies have examined interventions and network-evolution mechanisms related to polarization. For instance, Coscia and Rossi~\cite{coscia2022minimizing} model users who can unfollow sources and unfriend other users with sufficiently different opinions, and show that such conflict-avoidance behavior can increase polarization. Other approaches explicitly modify exposure patterns to encourage consensus~\cite{racz2023towards}. Other works have focused on selecting key users to limit misinformation, spread fact-checking information, or apply small changes to networks and user opinions to reduce polarization~\cite{tong2020stratlearner,ninomiya2025mitigating,zhang2025reducing}. In addition, Bolzern et al.~\cite{bolzern2020opinion} studied how algorithms can control user interactions, showing that even unbiased interventions may lead to polarization. However, these studies rely on simplified interaction models and do not capture language-based communication.

\paragraph{\textbf{LLM-Based Social Network Simulation}}
Recent work has examined LLM-driven social bots in online networks, showing that such agents can generate realistic interactions and influence communities through coordinated harmful behaviors~\cite{li2025understanding}.
Advances in LLMs have also enabled agent-based simulations in which agents interact through natural language, addressing key limitations of classical opinion dynamics models. Prior work shows that these agents can exchange messages, update beliefs, and reproduce phenomena such as consensus and polarization~\cite{chuang2024simulating,cau2025language,wang2025decoding,gu2025large}, with some models incorporating language-driven opinion updates and persuasion effects~\cite{cau2025language}.
Subsequent studies have extended these simulations to more complex settings, showing how recommender systems, content exposure, and agent interactions can shape echo chambers, polarization, and online discourse~\cite{bojic2025agent,tornberg2023simulating,wang2025decoding}.

\paragraph{\textbf{Opinion Manipulation and LLM Agents}}
Manipulation has been extensively studied in classical opinion dynamics. Chen and Rácz~\cite{chen2021adversarial} showed that an attacker can substantially increase disagreement and polarization by taking over a limited number of users and shifting their opinions to extreme values. Tu et al.~\cite{tu2023adversaries} further demonstrated that such attacks can remain effective even when the attacker only knows the network structure. More recent work has considered more strategic attackers. Zareer and Selmic~\cite{zareer2025maximizing} used reinforcement learning to learn manipulation strategies, whereas Jalan and Papachristou~\cite{jalan2026opinion} extended the setting to multiple strategic agents with possibly conflicting goals.

Recent work has also begun to explore manipulation in LLM-based multi-agent systems. Touzel et al.~\cite{touzel2024simulation} developed a social network simulation to study AI-driven manipulation and demonstrated how manipulators can influence voting behavior and election outcomes.
Qiao et al.~\cite{qiao2025botsim} introduced BotSim, an LLM-powered botnet simulation for studying disinformation spread and bot detection. Orlando et al.~\cite{orlando2025emergent} showed that groups of LLM agents can coordinate their posts and interactions to support a shared influence campaign, while Schroeder et al.~\cite{schroeder2026malicious} discussed how malicious AI swarms can coordinate at scale to manipulate public opinion.

However, existing work has not fully examined how adversarial behavior affects polarization and extremization in LLM-based social networks.

\section{Preliminaries}
\label{sec:Preliminaries}
This section introduces the graph-based opinion dynamics framework used in this study. First, the network structure and node attributes are defined. 
Then, the measures used to evaluate the impact of opinion manipulation and mitigation strategies are presented. Finally, the LLM-based communication process and opinion dynamics are discussed.

\subsection{Definitions}

A social network is modeled as a directed graph $G=(V,E)$, where each node $v\in V$ represents a user, and each edge $(v,u)\in E$ indicates that user $v$ can influence user $u$. Undirected graphs are treated as special cases by replacing each undirected edge with two directed edges in opposite directions.
For each node $v\in V$, the sets of outgoing and incoming neighbors are denoted by
$\Gamma_v^+=\{u\in V\mid (v,u)\in E\}$ and
$\Gamma_v^-=\{u\in V\mid (u,v)\in E\}$, respectively. The out-degree of $v$ is defined as
$\deg_v^+=|\Gamma_v^+|$.

Each node $v$ is associated with three attributes: an opinion $o_v^t \in [-1,1]$ at time step $t$, a stubbornness parameter $s_v \in [0,1]$, and an activeness parameter $a_v \in [0,1]$. The stubbornness parameter $s_v$ captures the resistance of node $v$ to opinion change, with larger values indicating lower susceptibility to external influence. The activeness parameter $a_v$ represents the probability that node $v$ generates content during the simulation.

Let $V=\{v_1,v_2,\dots,v_n\}$ denote the set of nodes in the graph. For notational simplicity, we define the opinion vector at time step $t$ as $\mathbf{o}^t=(o_{v_1}^t,o_{v_2}^t,\dots,o_{v_n}^t)^\top$. In particular, the initial opinion vector is
$\mathbf{o}^0=(o_{v_1}^0,o_{v_2}^0,\dots,o_{v_n}^0)^\top$.
The opinions evolve over time from the initial vector $\mathbf{o}^0$, while the stubbornness $\mathbf{s}=(s_{v_1},s_{v_2},\dots,s_{v_n})^\top$ and activeness $\mathbf{a}=(a_{v_1},a_{v_2},\dots,a_{v_n})^\top$ remain fixed throughout the simulation.

\subsection{Objective Measures}

The primary objective of this study is to investigate the effects of manipulator strategies and mitigation mechanisms on the global opinion state of the network. In particular, we focus on two complementary quantities:
\textbf{polarization} and \textbf{extremization}. These measures capture both opinion diversity across the network and overall opinion extremity.

The polarization at time step $t$ is defined as follows:

\begin{equation}
\mathcal{P}^t = \frac{1}{|V|}\sum_{v \in V}(o_v^t - \bar{o}^t)^2,
\quad
\bar{o}^t = \frac{1}{|V|} \sum_{v \in V}o_v^t,
\label{eq:polarization}
\end{equation}
where $\bar{o}^t$ denotes the average opinion at time step $t$. Polarization measures the variance in opinions across a network, with larger values indicating greater opinion diversity among users.

Extremization captures the overall extremity of opinions in the network and is computed as the average absolute opinion value across all users.
Larger values indicate that users' opinions are, on average, closer to the extremes $\pm 1$.
Extremization at time step $t$ is defined as follows:
\begin{equation}
\mathcal{E}^t =\frac{1}{|V|} \sum_{v \in V} |o_v^t|.
\label{eq:extremization}
\end{equation}

Unlike polarization, which depends on disagreement between users, extremization measures the overall intensity of opinions, regardless of whether users agree or disagree with one another. Together, these two measures provide a complementary view of collective opinion dynamics within the network.

\subsection{LLM-Based Opinion Dynamics}

Given a graph $G=(V,E)$ and node attributes, including initial opinions $\mathbf{o}^0$, stubbornness $\mathbf{s}$, and activeness $\mathbf{a}$, the system evolves over $T$ update steps. For $t=0,1,\dots,T-1$, the state at time $t$ is updated to produce the state at time $t+1$ through the following operations:
\begin{enumerate}
\item \textbf{Node activation.}
A node $v_{\text{act}} \in V$ is selected for activation. Specifically, each node $v \in V$ is selected with probability $\frac{a_v}{\sum_{u \in V} a_u}$.

\item \textbf{Content generation.}
The activated node $v_{\text{act}}$ generates a post using an LLM conditioned on its opinion $o_{v_{\text{act}}}^{t}$.

\item \textbf{Content propagation.}
The generated post is broadcast to the outgoing neighbors of $v_{\text{act}}$, defined as the recipient set $R(v_{\text{act}})=\Gamma^{+}(v_{\text{act}})$. Here, $R(\cdot)$ denotes the set of recipients of a message; in the basic communication model it coincides with the outgoing neighborhood $\Gamma^{+}(\cdot)$, although later interventions may modify the recipient set.

\item \textbf{Opinion update.}
Each recipient node $u \in R(v_{\text{act}})$ updates its opinion based on the received post, its previous opinion $o_u^{t}$, and its stubbornness $s_u$, resulting in an updated opinion $o_u^{t+1}$.

\item \textbf{State preservation.}
Nodes that do not receive the post retain their previous opinions, i.e.,
$o_u^{t+1} = o_u^{t}$ for all $u \in V \setminus R(v_{\text{act}})$.

\end{enumerate}

This procedure models repeated one-to-many language-based interactions in a social network, where opinions evolve through iterative content generation and exposure to neighbors' content.

\section{Opinion Manipulation}
\label{sec:manipulation}

In this section, we describe the opinion manipulation strategies considered in this study, along with different levels of manipulator persistence.
Opinion manipulation refers to a setting in which a limited set of manipulators $\mathcal{A}\subset V$, with $|\mathcal{A}|=k_A$, aims to influence other users to increase polarization and extremization in the network. The parameter $k_A$ denotes the manipulator budget.
In our framework, manipulators do not directly change the opinions of other users. Instead, they generate natural language content that gradually pushes exposed users toward more extreme views. We model manipulation through multiple factors: the number of manipulators, the strategy used to select them, and the changes applied to their configuration, including opinions, activeness, and stubbornness.

\subsection{Manipulator Selection Strategies}

We consider a broad spectrum of manipulation strategies, building on commonly used methods for social network control, including viral marketing and adversarial attacks. The strategies considered in this work are described below.


\begin{itemize}
    \item \textbf{Random strategy:} nodes are selected uniformly at random.
    \item \textbf{Out-degree-based strategy:} nodes with the highest out-degree are selected, targeting highly connected users.
    \item \textbf{Betweenness-based strategy:} nodes with the highest betweenness centrality~\cite{freeman1977set} are selected, targeting structurally important nodes that frequently lie on shortest paths between other nodes.
    \item \textbf{Community-aware greedy strategy:} communities are first ranked by a priority score, and manipulators are then selected greedily from high-priority communities using out-degree and opinion alignment. The details are as follows.
\end{itemize}

For the community-aware greedy strategy, we consider the following steps:
\begin{enumerate}

    \item We first apply Louvain community detection algorithm~\cite{blondel2008fast} to identify communities in the network.
    
    \item Each community $C_i$ is assigned a priority score $(1 - |\bar{o}_{C_i}^0|)\cdot |C_i|$, where $\bar{o}_{C_i}^0$ denotes the average opinion of the community, giving higher priority to larger and less extreme communities, which means communities with lower $|\bar{o}_{C_i}^0|$.
    
    \item Communities are then partitioned into positive and negative groups based on the sign of their average opinion, and manipulator allocation alternates between these groups whenever possible to influence both sides of the opinion spectrum.
    
    \item Within each selected community, candidate nodes are restricted to those whose opinion sign aligns with the community average ($o_v^0 \cdot \bar{o}_{C_i}^0 \ge 0$). This sign alignment ensures that manipulation pushes users further toward opinions they already hold, rather than reversing their stance. Among these candidates, nodes with higher out-degrees are prioritized, and in the case of ties, the node with smaller $|o_v^0|$ is selected.

\end{enumerate}

Among these strategies, the random strategy serves as a naive baseline, centrality-based strategies target high-impact nodes based on their structural positions, and the community-aware greedy strategy provides an alternative by selecting nodes from different communities and accounting for community structure.

\subsection{Manipulator Persistence Levels}

The persistence level specifies whether a selected manipulator maintains its extreme opinion during the simulation or remains susceptible to future influence. We consider two persistence levels for the selected manipulator set $\mathcal{A}$. In both settings, the selected nodes are first initialized with extreme opinions ($\pm 1$).

\begin{itemize}
    \item \textbf{Persistent manipulators:} selected nodes remain fixed at extreme opinions throughout the simulation.
    \item \textbf{Susceptible manipulators:} selected nodes are initialized with extreme opinions but can still update their opinions during the simulation.
\end{itemize}

For each selected manipulator $v \in \mathcal{A}$, its opinion is initialized according to its initial opinion as follows:
\[
o_v^0 \gets
\begin{cases}
\operatorname{sign}(o_v^0) & \text{if } o_v^0 \neq 0, \\
+1 \text{ or } -1 \text{ uniformly at random} & \text{otherwise.}
\end{cases}
\]

This design pushes nodes further toward opinions they already hold, better reflecting manipulation that amplifies existing leanings.
For this reason, in the community-aware greedy strategy, node selection is restricted to nodes whose opinion sign is aligned with the average opinion sign of their community, that is,
$o_v^0 \cdot \bar{o}_{C_i}^0 \ge 0.$
To increase their influence, the activeness and stubbornness of the selected manipulators are increased by parameter $\delta$. Specifically, for each selected node $v$,
$a_v \gets \min(1, a_v + \delta)$
and
$s_v \gets \min(1, s_v + \delta).$
We use $\delta = 0.5$ for susceptible manipulators and $\delta = 1$ for persistent manipulators. Consequently, manipulators become more active and less susceptible to the influence of other users. In the persistent setting, nodes become fully stubborn and therefore no longer update their opinions.

\section{Mitigations}
\label{sec:mitigation}

Mitigations are mechanisms designed to reduce polarization and limit the influence of manipulators. We consider two categories: \textbf{reactive mitigation} and \textbf{proactive mitigation}.
Reactive mitigations operate by selecting a set of users as moderators $\mathcal{M} \subset V \setminus \mathcal{A}$, with $|\mathcal{M}| = k_M$, and assigning them behaviors that counteract manipulative influence through their interactions.
Proactive mitigations apply general rules that shape how interactions occur across the network. These mechanisms are not tied to particular users, but instead modify exposure patterns, connectivity, content filtering, or activity dynamics.
For example, mechanisms may adjust feeds, introduce neutral content, recommend new connections, filter extreme content, or encourage broader participation across the network.

\subsection{Reactive Mitigations}

In the reactive setting, selected moderators $\mathcal{M} \subset V \setminus \mathcal{A}$, where $|\mathcal{M}| = k_M$, use language-based interactions to stabilize opinions and mitigate manipulative influence. Two types of reactive moderator behaviors are considered.

\begin{figure}[!htpb]
\centering
\includegraphics[width=0.5\linewidth]{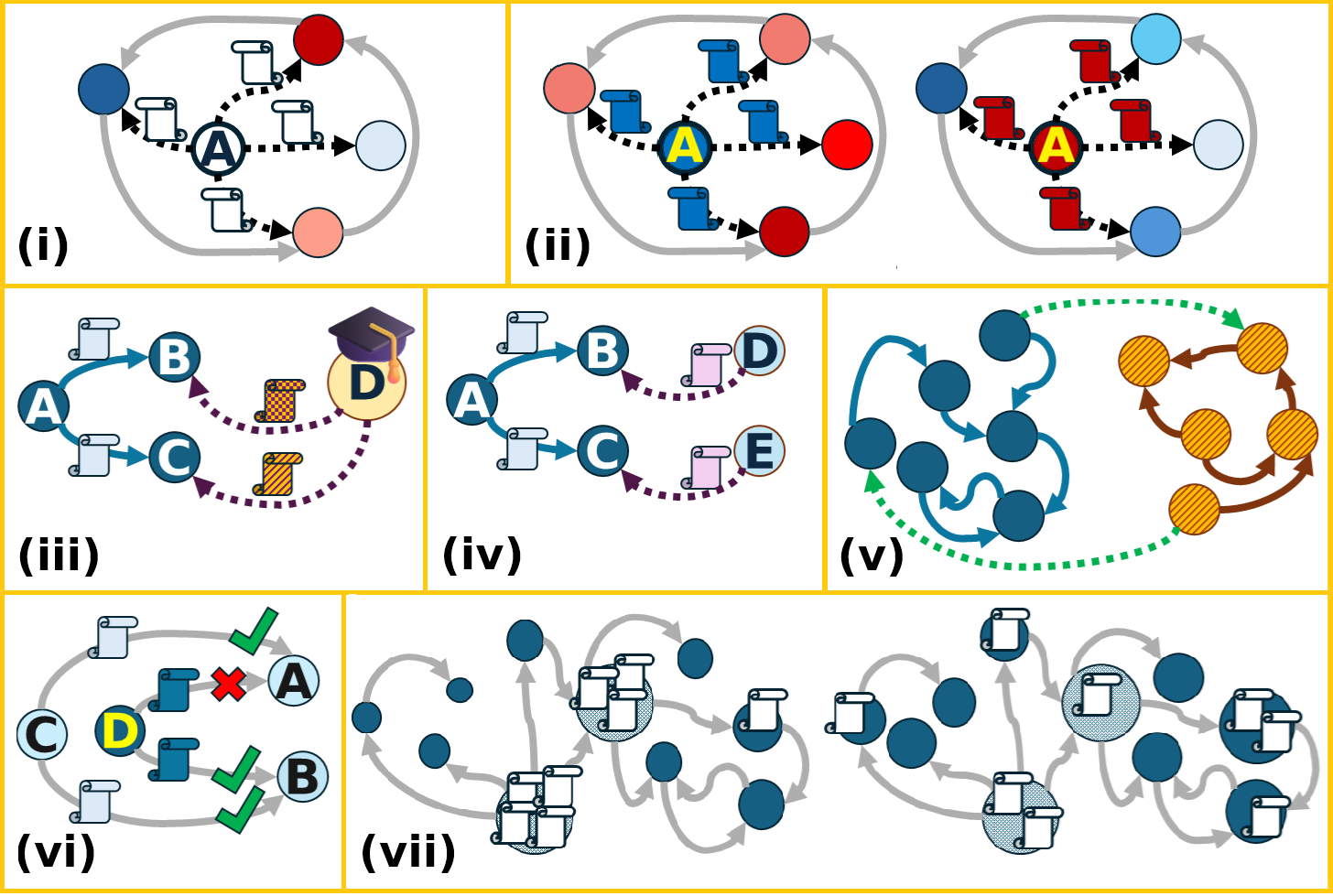}
\caption{Illustration of the reactive and proactive mitigations considered in this work. 
Plots (i) and (ii) show reactive mitigations, where node colors represent agent opinions and message colors represent content stances: blue and red indicate opposing viewpoints, darker shades indicate more extreme opinions or stances, and white denotes neutrality. 
Plots (iii)--(vii) show proactive mechanisms that modify exposure, connectivity, filtering, and activity patterns. 
In (iii)--(v), colors indicate content diversity; in (vi), darker blue indicates more extreme content and crossed-out messages are filtered; and in (vii), larger nodes represent more active agents that generate content more frequently.}
\label{fig:CounterMeasures}
\end{figure}

\paragraph{\textbf{Neutral Moderators}}
Neutral moderators maintain a neutral opinion throughout the simulation, that is,
$o_v^t = 0$
for all $t$. The generated content reflects a neutral stance and aims to pull neighbors' opinions toward the neutral opinions. This behavior corresponds to fact-checkers or users who try to share accurate information and reduce conflict in discussions. In Figure~\ref{fig:CounterMeasures}, plot \texttt{(i)} shows that node \texttt{A} shares neutral content with its neighbors regardless of their opinions.

\paragraph{\textbf{Contrarian Moderators}}
Contrarian moderators adapt their opinions according to the local neighborhood. Specifically, at each time step,
$o_v^t = -\operatorname{sign}\left(\sum_{u \in \Gamma_v^+} o_u^{t-1}\right),$
and
$o_v^t = 0$
if the sum is zero. Consequently, moderators react in the opposite direction to the dominant local opinion to counter reinforcement and polarization. This behavior can model interventions that introduce communities to different opinions to reduce polarization. Figure~\ref{fig:CounterMeasures}, plot \texttt{(ii)}, illustrates a contrarian moderator taking the opposite stance to the average opinion of its neighbors (blue and red denote opposing opinions). 

\subsubsection*{\textbf{Moderator Selection Strategies}}

We define a \textbf{moderator pool}
$\mathcal{M}_{\text{pool}} \subseteq V \setminus \mathcal{A}$
representing controllable or cooperative nodes, such as volunteers, trusted users, or fact-checkers. From this pool, a moderator budget of
$k_M$
nodes is selected to form the moderator set
$\mathcal{M} \subseteq \mathcal{M}_{\text{pool}}.$
This setting reflects practical scenarios in which only a subset of users is willing or authorized to moderate content and interactions in the network.

Moderator selection is performed independently of the manipulators. Specifically, we select
$k_M' = \min(k_M, |\mathcal{M}_{\text{pool}}|)$
nodes from the moderator pool to form $\mathcal{M}$.
The same selection strategies used for manipulators, namely \textbf{random}, \textbf{out-degree-based}, \textbf{betweenness-based}, and \textbf{community-aware greedy} strategies, are also considered for moderator selection. However, moderator selection is restricted to nodes within the \textit{moderator pool} $\mathcal{M}_{\text{pool}}$. For the \textit{community-aware greedy strategy}, we use a modified procedure: after detecting communities with the Louvain algorithm, only nodes in $\mathcal{M}_{\text{pool}}$ are considered as candidates within each community. 

The communities are processed in decreasing order of the number of available candidates. Within each community, candidates are ranked by out-degree, and the highest-ranked nodes are selected until the moderator budget $k_M'$ is reached. This strategy prioritizes more connected moderator candidates while maintaining coverage across communities.

Once selected, moderators participate in the same interaction process as other nodes while following their assigned behavior. Neutral moderators maintain fixed neutral opinions, whereas contrarian moderators adapt their opinions to oppose dominant local trends. For all moderator nodes $v \in \mathcal{M}$, the activeness is set to $a_v = 1$.

\subsection{Proactive Mitigations}

In addition to reactive mitigations, we consider several proactive mitigations to improve the resistance to polarization and extremization before manipulative effects. Unlike reactive approaches, these operate at the system level and do not rely on the selection or control of specific nodes. Instead, they modify interaction patterns, exposure, or node activity to reduce the amplification of similar opinions or increase exposure to diverse opinions. The following mechanisms are considered:

\paragraph{\textbf{Neutral-Content Exposure}}

During opinion updates, nodes may observe externally generated neutral-content posts. This introduces balanced information into the interaction process and encourages moderation of opinions. 
In practice, this can represent fact-checking labels or neutral news recommendations in user feeds.
Figure~\ref{fig:CounterMeasures}, plot \texttt{(iii)}, illustrates neutral-content exposure, where nodes \texttt{B} and \texttt{C} receive additional neutral information from trusted source \texttt{D} along with the content propagated by node \texttt{A}.

\paragraph{\textbf{Feed-Based Exposure}}

During opinion updates, nodes may observe randomly selected posts generated earlier in the simulation. This exposes users to information beyond their immediate neighbors and reduces the amplification of similar opinions. Such a mechanism may correspond to diversified recommendation feeds or algorithms that expose users to a broader range of content. In Figure~\ref{fig:CounterMeasures}, plot \texttt{(iv)}, nodes \texttt{B} and \texttt{C} receive content from node \texttt{A}, while also being exposed to additional earlier posts from other nodes.

\paragraph{\textbf{Broadening Social Ties}}

For each node, $k$ additional incoming edges are randomly added from previously unconnected nodes. This increases exposure to diverse opinions and weakens structural isolation between communities.
Such mechanisms may correspond to recommendation systems that encourage interactions with users holding different opinions or users from outside their local communities. In Figure~\ref{fig:CounterMeasures}, plot \texttt{(v)} shows that the green edges represent additional cross-community connections between previously unrelated nodes (nodes belonging to different communities are shown in different colors). 

\paragraph{\textbf{Extreme-Content Filtering}}

When highly extreme nodes generate content, their spread to neighbors is reduced with a certain probability, limiting the spread of extreme opinions. Specifically, for each receiving node, content generated by a node with
$|o_v| > 0.5$
is ignored with probability $0.5$ during the opinion update process. This mechanism can correspond to platform-level moderation policies that reduce the visibility of highly extreme or polarizing content by filtering them, or to users naturally ignoring strongly biased viewpoints. In Figure~\ref{fig:CounterMeasures}, plot \texttt{(vi)}, node \texttt{D} represents a highly extreme node, whereas node \texttt{C} is less extreme. As illustrated, the content generated by node \texttt{D} is rejected by node \texttt{A}, whereas the content generated by node \texttt{C} is received by all neighbors.

\paragraph{\textbf{Activeness Boosting}}

Node activeness values are globally increased by a small amount
$\delta_{\text{boost}} \in [0,1]$.
Specifically, for each node $v \in V$, we update
$a_v = \min(a_v+\delta_{\text{boost}}, 1).$
This reduces the relative dominance of a small number of highly active users and promotes broader participation across the network. In practice, this may correspond to platform mechanisms that encourage participation from less active users through content promotion and engagement incentives. Figure~\ref{fig:CounterMeasures}, plot \texttt{(vii)}, illustrates how increasing activeness across the network reduces the dominance of a few highly active nodes (node diameter represents activeness).

These mitigation mechanisms are lightweight and broadly applicable, and can be viewed as conceptualized versions of real-world instances such as recommendation systems, content exposure policies, and moderation mechanisms.

\section{Experiments}
\label{sec:experiments}

In this section, we first describe the experimental setup, including the datasets, initialization process, and LLM configuration used in the simulations. We then evaluate the vulnerability of social networks to opinion manipulation attacks and analyze the effectiveness of different 
reactive and proactive mitigation strategies.

\subsection{Experimental Setup}

The following paragraphs describe the datasets, initialization process, community detection procedure, and LLM configuration used in the experiments.

\paragraph{\textbf{Graph Models and Datasets}}
The framework was evaluated on both synthetic and real-world social networks. Synthetic graphs are generated using Hyperbolic Random Graphs (HRG)~\cite{krioukov2010hyperbolic}, which capture important structural properties such as clustering and community structure. For real-world evaluation, social network datasets from Twitter and Reddit are used, where node opinions are provided~\cite{chitra2019understanding}.
All graphs were treated as directed; undirected graphs were converted into directed graphs by replacing each undirected edge with two directed edges in opposite directions.

\paragraph{\textbf{Initialization of Node Attributes}}

For synthetic graphs, opinions were initialized using a community-aware process. Communities were first identified using the Louvain community detection algorithm. Because Louvain is defined for undirected graphs, community detection is performed on an undirected version of the network obtained by ignoring edge directions. Each community $C_i$ is then assigned a mean opinion $o_{C_i} \sim \mathcal{N}(0,\sigma_1^2)$, and the initial opinion of each node $v \in C_i$ is sampled from $o_v^0 \sim \mathcal{N}(o_{C_i}, \sigma_2^2)$. This procedure produces locally correlated opinions while preserving variation among individuals within the same community.

To capture heterogeneous user behavior commonly observed in opinion dynamics studies, each node is assigned a stubbornness value and an activeness value. Stubbornness was sampled from a normal distribution with mean $0.5$, following prior work~\cite{meylahn2024opinion}, whereas activeness was sampled from a long-tailed distribution to reflect the unequal participation levels typically observed in social networks~\cite{barabasi2005origin}.

Finally, opinions were clipped to $[-1,1]$, and stubbornness and activeness values were clipped to $[0,1]$. In addition, for real-world datasets, opinions were rescaled to the interval $[-1,1]$ to ensure consistency across experiments.

\paragraph{\textbf{LLM Configuration and Computational Setup}}

The experiments used OpenAI models accessed via API~\cite{openai2024api}, primarily \texttt{GPT-4.1-mini}, with additional experiments conducted using \texttt{GPT-4o-mini} and the DeepSeek API~\cite{bi2024deepseek}. The temperature parameter was set to $0.7$ for post-generation and $0.2$ for opinion reporting. To prevent context leakage between agents, each user is assigned an independent LLM session. Prompt templates were intentionally kept minimal and structured to encourage consistent behavior across simulations. The full prompt templates and additional details are provided in Appendix~\ref{APDX:prompts}.
All experiments used CPU-only resources, with each job allocated two CPU cores and 1GB of RAM. Since language generation was performed via external APIs, local resources were mainly used for simulation and data processing.

\paragraph{\textbf{Baseline and Reporting}}
All reported results are averaged over 10 independent runs. In each experiment, the baseline is computed on the same graph without any manipulation or mitigation. Unless otherwise stated, the manipulation setting is persistent, and manipulators are selected using the community-aware greedy strategy with $k_A = 6$.

\subsection{Vulnerability to Opinion Manipulation}

We analyze how manipulators affect polarization and extremization under different selection strategies, persistence settings, and budgets across various networks.

\paragraph{\textbf{Manipulator Selection Strategies}}

\begin{figure}[!htpb]
\centering
\includegraphics[width=0.7\linewidth]{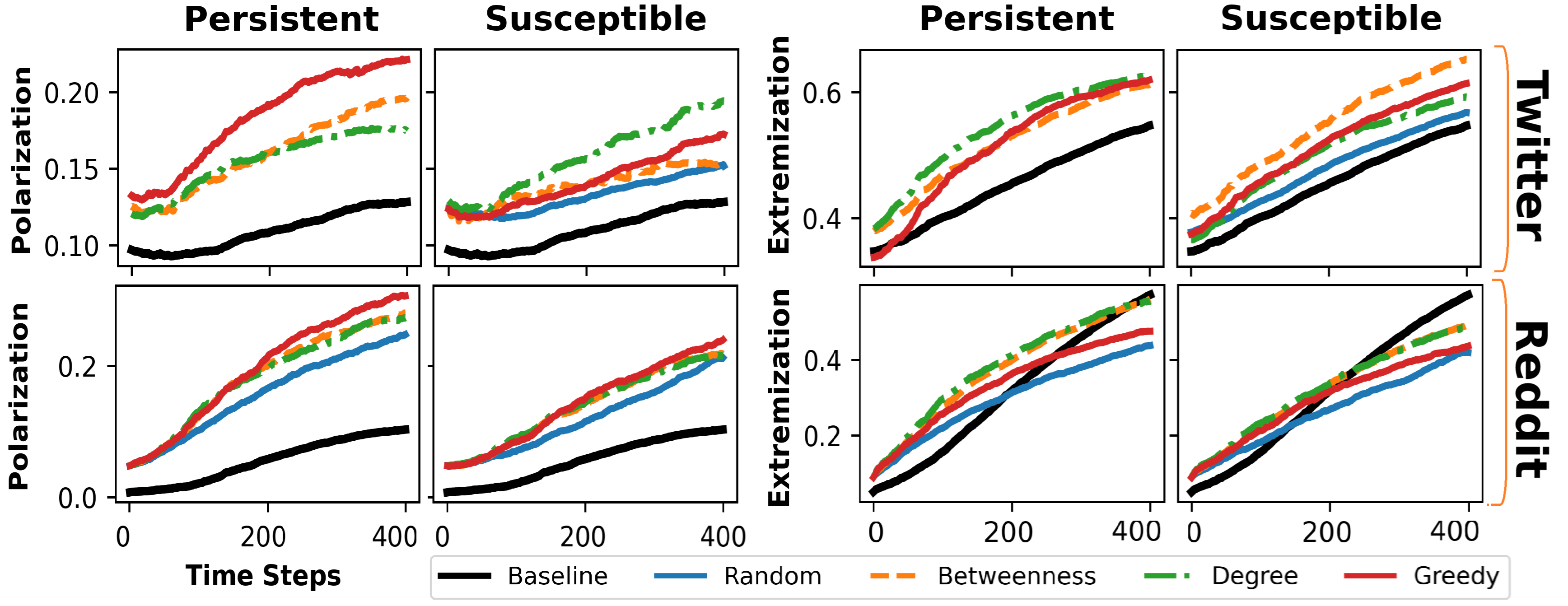}
\caption{Manipulator selection strategies on Reddit and Twitter graphs under susceptible and persistent manipulators. Different lines represent different manipulator selection strategies.}
\label{fig:NodeSelectionStrategies}
\end{figure}

The effectiveness of different manipulator selection strategies, including random, out-degree-based, betweenness-based, and community-aware greedy strategies, is compared to understand how manipulator placement affects the spread of extreme opinions and polarization.
Figure~\ref{fig:NodeSelectionStrategies} shows the evolution of polarization over time under different manipulator selection strategies. Each colored line represents the normalized increase in polarization compared to the baseline. The community-aware greedy strategy consistently leads to the highest polarization, suggesting that distributing manipulators across communities increases their overall influence. Overall, the choice of selection strategy plays a key role, with community-aware methods being the most effective, while random and standard centrality-based strategies are less effective in increasing polarization and extremization.

\paragraph{\textbf{Susceptible versus Persistent Manipulators}}
As shown in Figure~\ref{fig:NodeSelectionStrategies}, the effects of susceptible and persistent manipulator settings were evaluated. Persistent manipulators produced much higher polarization, whereas susceptible manipulators often moved closer to the baseline behavior over time owing to their continued susceptibility to influence.

The results demonstrate that persistent manipulators induce much higher polarization and extremization than susceptible manipulators, highlighting the importance of highly active and persistent manipulators in amplifying opinion divergence.

\begin{figure}[!htpb]
\centering
\includegraphics[width=0.6\linewidth]{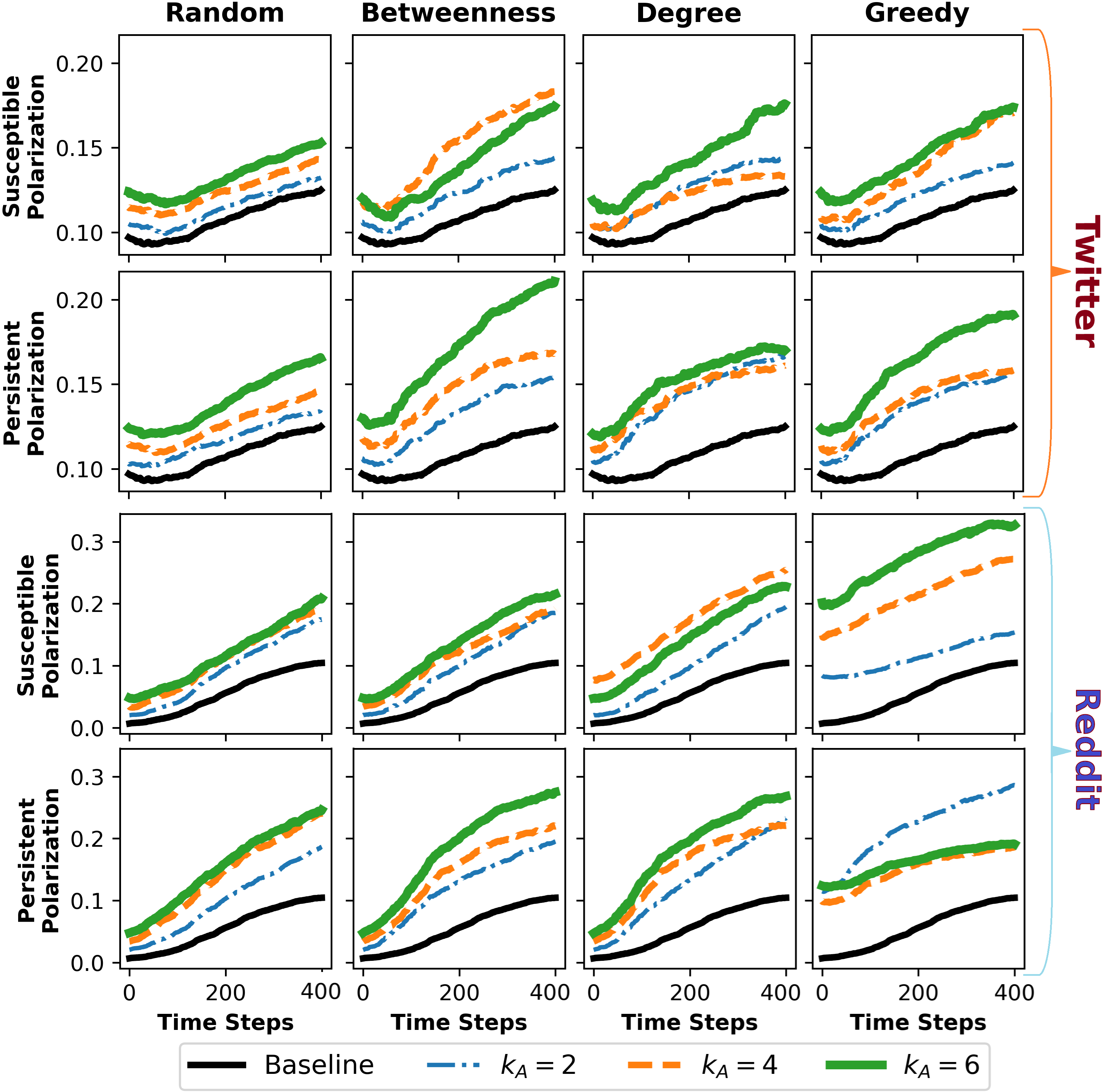}
\caption{Effect of changing the manipulator budget. The black line represents the baseline and the other lines correspond to different manipulator budgets $k_A \in \{2,4,6\}$. Each column represents a different node selection strategy.}
\label{fig:Ks}
\end{figure}

\paragraph{\textbf{Effect of Manipulator Budget}}

The sensitivity of the system to different manipulator budgets was further investigated. The experiments examined whether small manipulator groups were sufficient to significantly alter global opinion dynamics and whether the effect scaled proportionally with the manipulator budget.
Figure~\ref{fig:Ks} illustrates the effect of different manipulator budgets under multiple experimental settings. Across all cases, increasing the manipulator budget consistently accelerates polarization growth and leads to much higher final polarization levels than the baseline.
Based on these results, opinion manipulation attacks consistently increase polarization and extremization compared to the baseline. Moreover, as the manipulator budget increases, polarization tends to emerge more rapidly and reach higher levels.

\subsection{Reactive Mitigations}

Reactive mitigation strategies for reducing manipulative influence during interactions were evaluated. Specifically, we study neutral and contrarian moderators, as well as reactive moderator selection strategies under both normal and manipulative settings.

\begin{figure}[!htpb]
\centering
\includegraphics[width=0.5\linewidth]{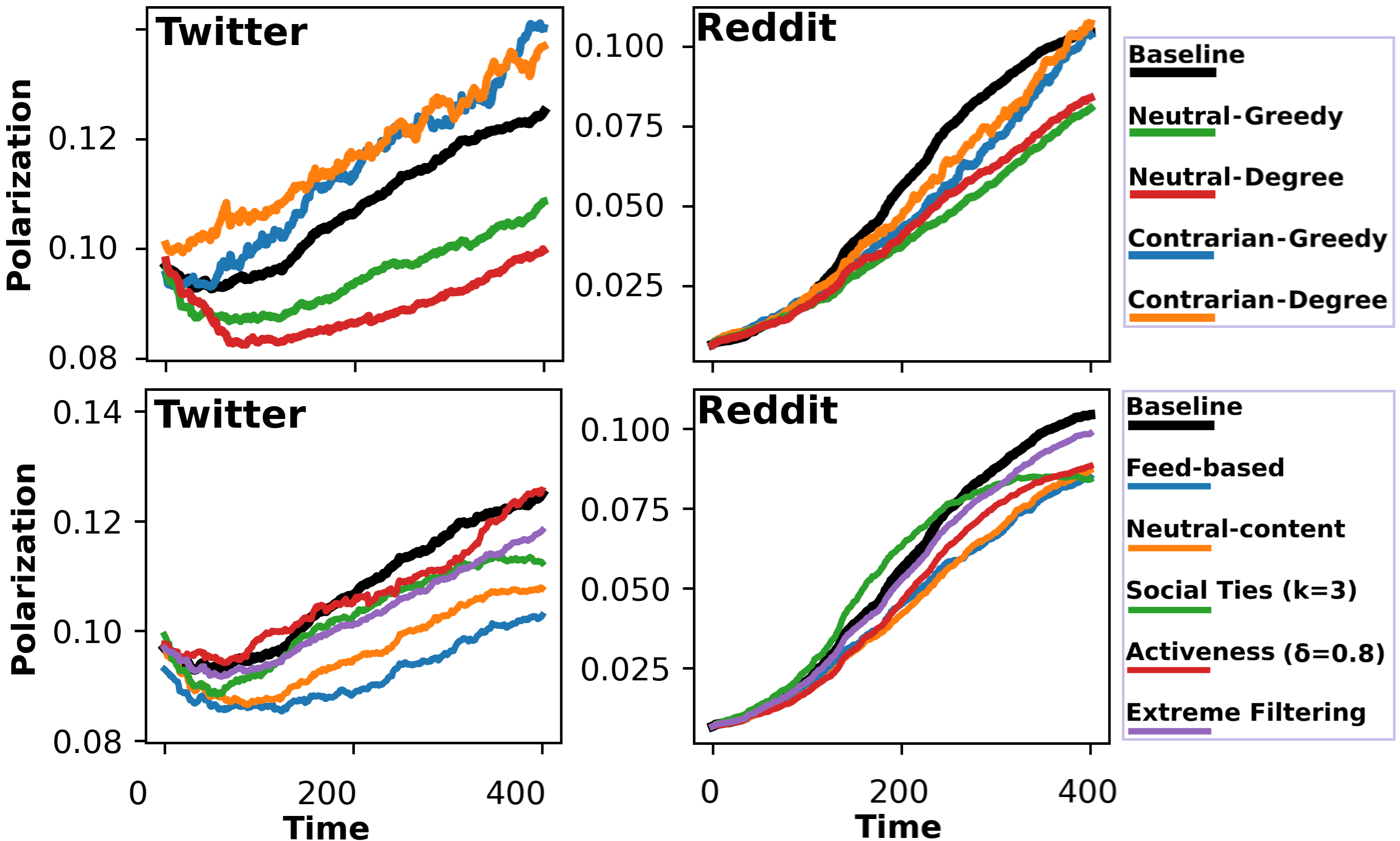}
\caption{Reactive and proactive mitigations without any opinion manipulation ($k_A = 0$).}
\label{fig:onlyCA}
\end{figure}

\paragraph{\textbf{Reactive Mitigations without Manipulators}}
The effect of reactive moderators was first analyzed without manipulators. The goal was to determine whether moderator-based approaches alone could reduce polarization and stabilize opinion dynamics. Neutral and contrarian behaviors were compared using different selection strategies and moderator budgets. As shown in Figure~\ref{fig:onlyCA} first row, introducing reactive moderators generally reduces polarization, whereas contrarian moderators are less effective, as they can increase polarization by introducing opposing extreme opinions, especially in the absence of manipulators.

\paragraph{\textbf{Mitigations under Opinion Manipulation}}

The effectiveness of reactive mitigations under manipulation was evaluated. We analyzed the interaction between manipulators and moderators under different persistence settings and budgets. Neutral and contrarian moderator behaviors are compared, along with different moderator selection strategies, to measure their impacts on polarization and extremization.
As shown in Figure~\ref{fig:combined-weak_vs_strong}, reactive moderators generally reduce polarization and slow the spread of extreme opinions. In highly polarized settings, contrarian moderators are generally more effective than neutral moderators, as they respond in the opposite direction to dominant local opinions. This suggests that contrarian reactive mitigation can be more effective in the presence of manipulators. However, in their absence, contrarian moderators may play a manipulative role by exposing less extreme users to polarized content.

\subsection{Proactive Mitigations}

The effectiveness of proactive mitigations without manipulators and in the presence of manipulators is evaluated.

\paragraph{\textbf{Mitigations without Manipulators}}

The isolated effect of mitigation mechanisms was first analyzed in the absence of opinion manipulation attacks. The evaluated mechanisms included neutral and contrarian moderators, broadening social ties, activeness boosting, feed-based exposure, neutral-content exposure, and extreme-content filtering. These experiments examined how each mechanism affects opinion dynamics without adversarial manipulation. As shown in the second row of Figure~\ref{fig:onlyCA}, in the absence of manipulators, neutral moderators do not substantially change the outcomes, since they are designed mainly to make the social network more robust. In contrast, sharing earlier random feeds as neutral content appears to be more effective, likely because this intervention directly changes the content users are exposed to.

\paragraph{\textbf{Mitigations under Opinion Manipulation}}

\begin{figure*}[!htbp]
\centering
\includegraphics[width=1\linewidth]{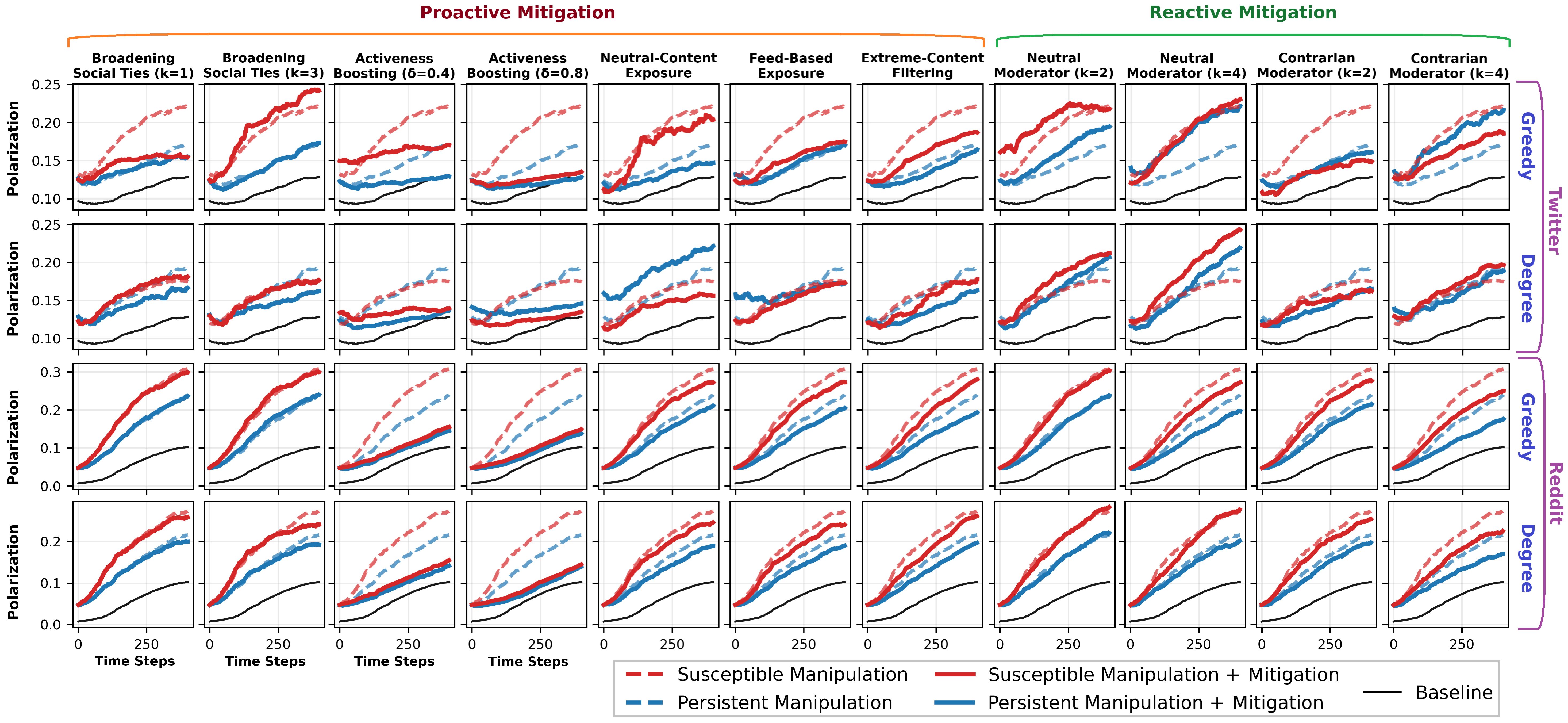}
\caption{Susceptible versus persistent manipulators in existence of different mitigations.}
\label{fig:combined-weak_vs_strong}
\end{figure*}

The robustness of proactive mitigations under manipulative settings was evaluated. In particular, we analyze whether proactive mechanisms can reduce the amplification of polarization caused by manipulators.
Figure~\ref{fig:combined-weak_vs_strong} shows that proactive mitigations reduce the impact of manipulators and improve robustness compared to the unmitigated setting. Overall, these results suggest that mitigation effects remain limited, with polarization levels staying higher than baseline. 
Activeness boosting has a notable effect. When all nodes become more active, the intervention counteracts manipulators by preventing them from becoming the dominant voice in the network.

\subsection{Robustness and Sensitivity Analysis}

The robustness of the proposed framework was evaluated under different simulation settings, language models, graph structures, and discussion environments.

\paragraph{\textbf{Sensitivity to Graph Families, Datasets, and Discussion Topics}}

\begin{figure}[!htpb]
\centering
\includegraphics[width=0.6\linewidth]{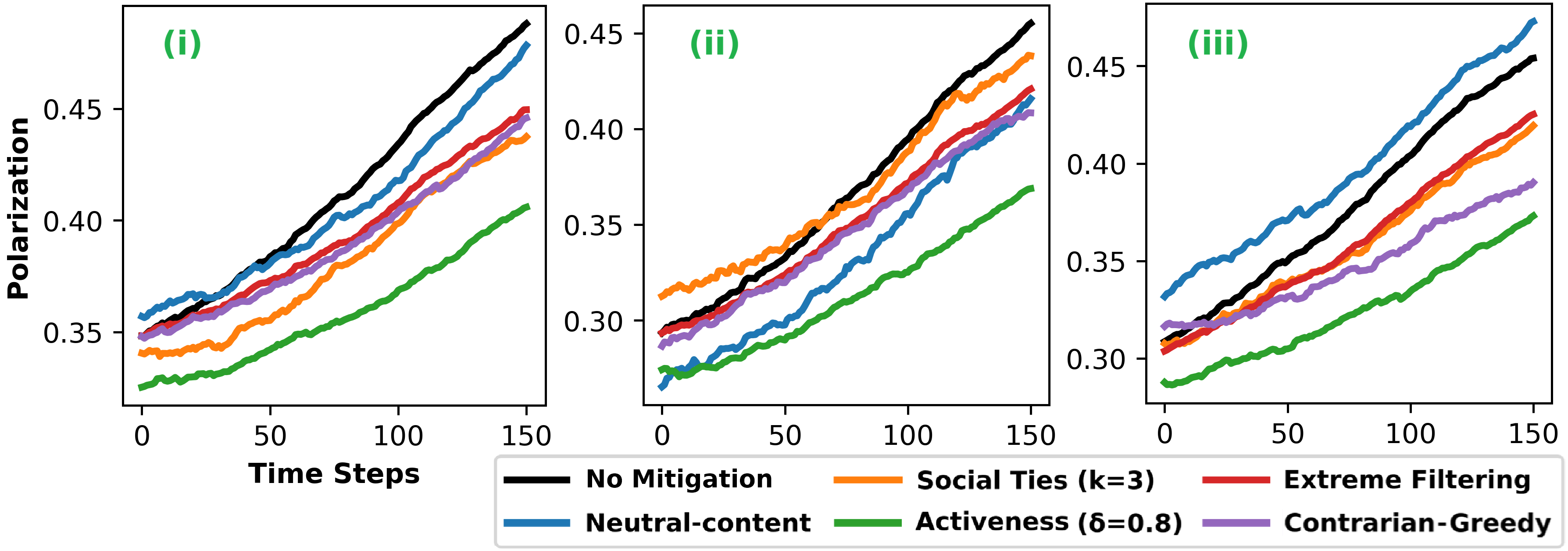}
\caption{HRG graph under three different discussion topics: (i) online learning, (ii) remote working, and (iii) universal basic income. 
}
\label{fig:topics}
\end{figure}

We studied how graph structure and discussion topics affect manipulation and mitigation performance. Earlier results were presented for two real-world networks: Reddit and Twitter. Here, we include an additional synthetic HRG graph under three topics: online learning, remote working, and universal basic income, representing discussions with different levels of social relevance and controversy (Figure~\ref{fig:topics}).
The results show that manipulative influence is stronger in networks with stronger community structure and stronger local opinion alignment, whereas mitigation strategies remain effective across graph types. Although different topics lead to different baseline levels of polarization and extremization, the relative behavior of manipulation and mitigation strategies remains largely consistent across settings.

\begin{figure}[!htpb]
\centering
\includegraphics[width=0.7\linewidth]{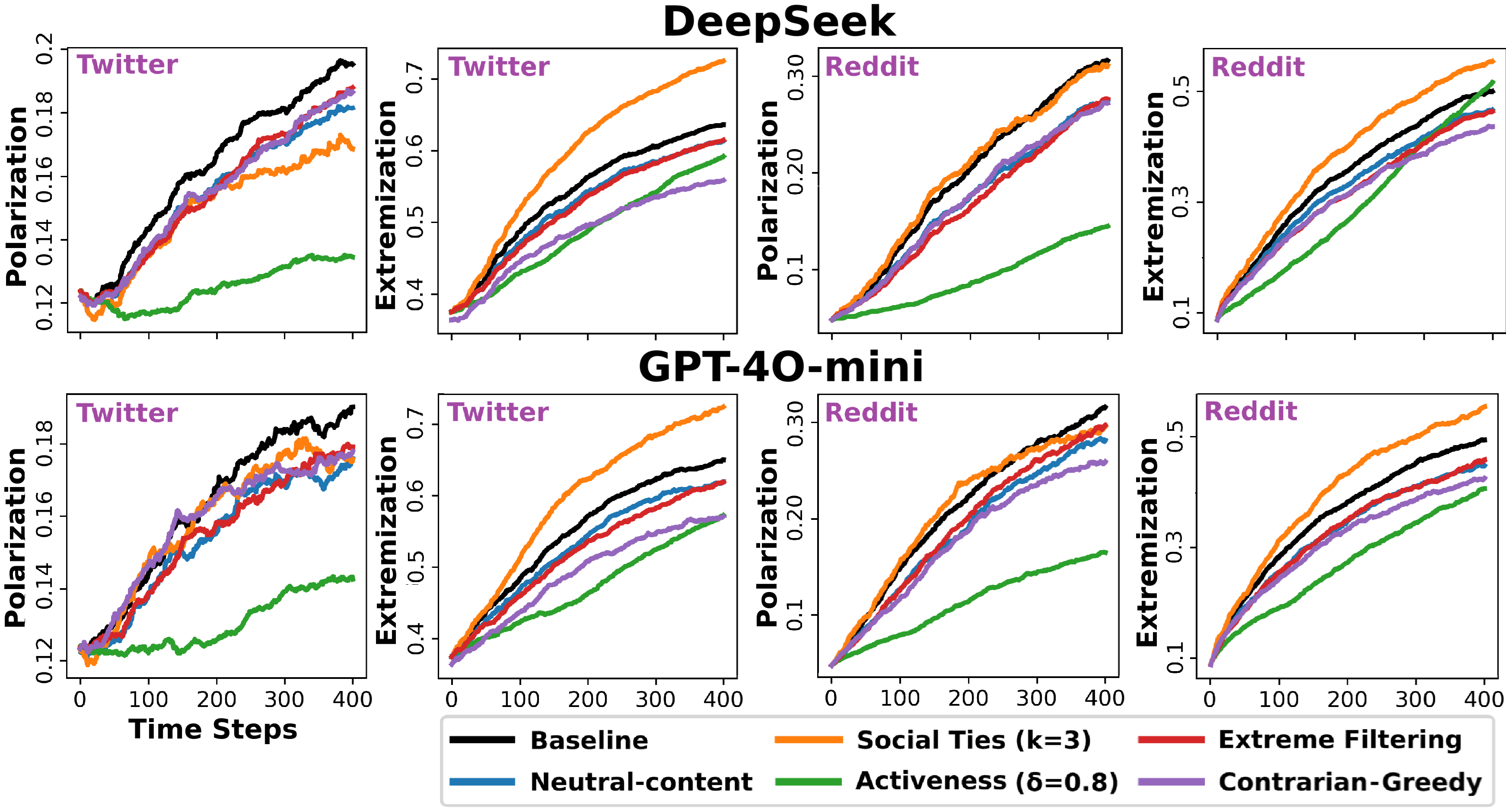}
\caption{Comparison across different LLMs. The first row corresponds to DeepSeek and the second row corresponds to GPT-4o-mini.}
\label{fig:DS}
\end{figure}

\paragraph{\textbf{Sensitivity to LLMs}}
The consistency of the observed opinion dynamics across different language models, including \texttt{GPT-4.1-mini}, \texttt{GPT-4o-mini}, and \texttt{DeepSeek}, was also evaluated. We examined whether the manipulative effects and mitigation performance remained stable across models with different generation behaviors. As shown in Figure~\ref{fig:DS}, the overall trends are qualitatively consistent, although the magnitude and rate of polarization growth vary across models.

\begin{figure}[!htpb]
\centering
\includegraphics[width=0.35\linewidth]{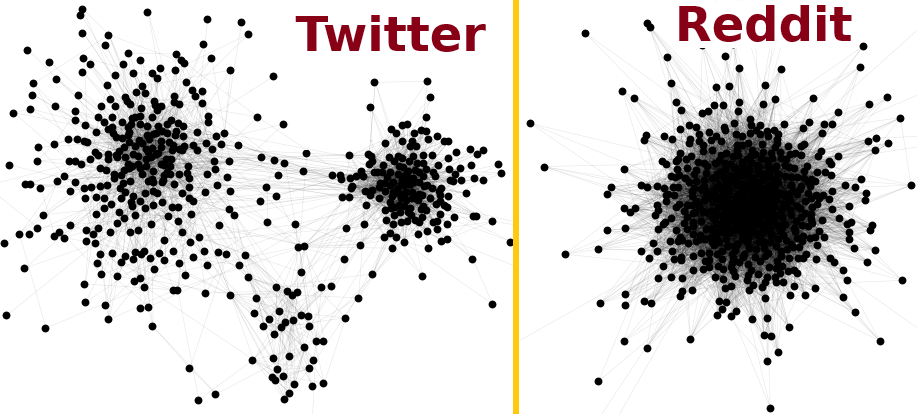}
\caption{Comparison of the Reddit and Twitter real-world datasets.}
\label{fig:redditvstwitter}
\end{figure}

\paragraph{\textbf{Reddit versus Twitter}}
An interesting observation is the difference in how the same models behave on the two datasets. Reddit is more likely to shift toward almost positive or almost negative states, and different moderators and manipulators produce more diverse behaviors on this dataset. This is reflected in the greater variety of trends observed on Reddit compared to Twitter.
One possible explanation is the difference in graph structure. The Twitter dataset appears to have more realistic social-network characteristics, with clearer community structure, whereas Reddit is closer to a single giant community, as shown in Figure~\ref{fig:redditvstwitter}. In real-world social networks, clearer communities may increase vulnerability because manipulators can target specific communities more effectively, and isolated communities may be easier to influence. Therefore, the structural differences between the two graphs can partly explain the differences in observed behavior.
\textit{Given these differences, observing similar trends across both datasets provides stronger validation for our findings}.

\section{Conclusion, Limitations, and Future Work}
\label{sec:conclusion}

We summarize our main findings and discuss the key limitations and future directions.

\paragraph{\textbf{Conclusion}}

In this study, we investigated adversarial opinion manipulation in social networks using an LLM-based simulation framework that captures language-based interactions. Our results show that social networks are highly vulnerable to manipulation; even a small number of adversarial manipulators can noticeably increase polarization and extremization. This effect depends on the selection of the manipulators, their persistence, and the number of manipulators used.

We found that both reactive and proactive mitigation strategies can reduce polarization compared to the unmitigated attack setting. However, neither approach can fully counter manipulator influence or restore the system to its no-manipulation state. These results highlight the limited effectiveness of the considered interventions in the face of adversarial influence.

\paragraph{\textbf{Limitations and Future Work}}

Our framework assumes a fixed network and does not model evolving connections. While this isolates manipulation effects, extending to dynamic processes such as link formation, edge removal, and evolving communities would enable a more realistic study of co-evolving manipulation and mitigation.
The interventions considered are simplified versions of real-world moderation. In practice, platforms face delayed detection, limited visibility, and adaptive adversaries. Future work should explore more realistic, adaptive strategies, including predictive interventions, flexible policies, and more strategic manipulators.
Finally, computational constraints limit simulation scale relative to real systems and restrict richer components such as recommendation algorithms and complex interactions. Scaling the framework and validating results with real-world data are key directions for future work.

\bibliographystyle{unsrtnat}
\bibliography{references}

\appendix
\section{Prompt Templates}
\label{APDX:prompts}

This appendix presents the prompt templates used in simulations. Two prompts are used: (i) a post-generation prompt for social-media-style posts and (ii) a response-generation prompt for updating user opinions after observing content. Prompts for both GPT-4o-mini and GPT-4.1-mini are provided below.
Prompt design can affect LLM-based simulations, and small changes may influence results. To reduce unrealistic outputs, we kept prompts simple and explicit. As a sanity check, we verified that a clique of extremely positive users did not rapidly shift to neutral or negative opinions, and vice versa.

\subsection{Post-Generation Prompt}

The post-generation prompt conditions text on the discussion topic and opinion intensity. The model generates a short post reflecting the given stance. We also experimented with multiple formulations for representing opinions, and found that numerical values in the range $[0,100]$ yield the most stable and consistent behavior across simulations.

\begin{lstlisting}[style=promptstyle]
You are a social network user.

Topic: Supporting {topic}
Your Opinion strength: {intensity_level}, which is equal to {percent} percent (0=against, 100=favor).

Write 1-3 sentences expressing your views.
Match tone to strength:
Low/moderate: casual or uncertain
High: decisive or strong
  
Do not argue for both sides.
Do not mention numbers or being an AI.
Post:
\end{lstlisting}

Samples generated by GPT-4.1-mini for ``Remote working'':

\begin{itemize}
\item \textbf{Opinion = -0.9}:
Remote work hurts collaboration and team culture. Most jobs work better in person.
\item \textbf{Opinion = 0}:
Remote work has benefits, but it also creates communication challenges. A hybrid model seems reasonable.
\item \textbf{Opinion = +0.9}:
Remote work gives people more flexibility and improves work-life balance. Companies should support it more.
\end{itemize}

\subsection{Response-Generation Prompt}

The response-generation prompt produces an updated opinion after exposure to a post. The response is conditioned on the agent's prior opinion and stubbornness.

\begin{lstlisting}[style=promptstyle]
You are scrolling social network.
Topic: {topic}
Your initial opinion strength: {intensity_level}, which is equal to {percent} percent (0=against, 100=favor).
Your stubbornness: {stubbornness} (higher = less change)
You see:
""" Supporting {post} """

Higher levels of stubbornness limit movement.
This is a common occurrence.
Return your new opinion (0-100).
Answer with ONE number only.
Answer:
\end{lstlisting}

\end{document}